\documentclass[pra,preprintnumbers,amsmath,amssymb,floatfix,superscriptaddress,aps,twocolumn,10pt]{revtex4-1}
\usepackage{graphicx}

\usepackage{xcolor}
\definecolor{applegreen}{rgb}{0.55, 0.71, 0.0}

\usepackage{hhline}
\usepackage{hyperref}
\usepackage{bm}
\usepackage{graphicx}
\usepackage{color}
\usepackage{subfigure}
\usepackage[binary,amssymb]{SIunits}
\DeclareGraphicsExtensions{.pdf,.eps,.png,.jpg,.mps}

\usepackage{color}
\definecolor{Blue}{rgb}{0.3,0.3,0.9}
\definecolor{Red}{rgb}{0.9,0.3,0.3}
\definecolor{Green}{rgb}{0.3,0.6,0.3}

\newcommand\doilink[1]{\href{http://dx.doi.org/#1}{#1}}

\renewcommand*\url[1]{\href{#1}{\texttt{#1}}}

\begin{document}

\title{Effects of random vacancies on the spin-dependent thermoelectric properties of silicene nanoribbon}

\date{\today} 

\author{D. Zambrano}
\affiliation{Departamento de F\'{i}sica, Universidad T\'{e}cnica Federico Santa Mar\'{i}a, Casilla Postal 110V, Valpara\'{i}so, Chile.}
\author{C. D. N\'u\~nez}
\affiliation{Departamento de F\'{i}sica, Universidad T\'{e}cnica Federico Santa Mar\'{i}a, Casilla Postal 110V, Valpara\'{i}so, Chile.}
\author{P. A. Orellana}
\affiliation{Departamento de F\'{i}sica, Universidad T\'{e}cnica Federico Santa Mar\'{i}a, Casilla Postal 110V, Valpara\'{i}so, Chile.}
\author{J. P. Ramos-Andrade}
\affiliation{Departamento de F\'isica, Universidad de Antofagasta, Av. Angamos 601, Casilla 170, Antofagasta, Chile.}
\author{L. Rosales}
\affiliation{Departamento de F\'{i}sica, Universidad T\'{e}cnica Federico Santa Mar\'{i}a, Casilla Postal 110V, Valpara\'{i}so, Chile.}

\begin{abstract}

The spin-dependent thermoelectric properties of silicene nanoribbon heterostructures are investigated, in which the central conductor contains a random distribution of vacancies and is connected to two pristine leads of the same material, placed in proximity to ferromagnetic insulators. The magnetic moments of the leads are analyzed in both parallel and antiparallel configurations. A tight-binding Hamiltonian and the Green’s function formalism are employed to calculate the spin-resolved thermoelectric properties of the system as functions of geometrical confinement and vacancy concentration. The results demonstrate an enhancement in charge and spin-dependent thermopower, resulting in an
improved thermoelectric efficiency at room temperature, which overcomes the limitations imposed by the classical Wiedemann–Franz law. These findings indicate that defective silicene nanoribbons are promising platforms for the development of efficient thermoelectric and spin-caloritronic devices.

\end{abstract}

\maketitle

\section{Introduction\label{sec:intro}}

The efficiency of thermal devices in converting heat into electricity, and vice versa, is quantified by thermoelectric efficiency, which is expressed through the figure of merit $ZT=S^2\mathcal{G} T/\kappa$ \cite{Goldsmid10}. Here, $S$ denotes the Seebeck coefficient, $\mathcal{G}$ represents the electronic conductance, and $\kappa$ refers to the total thermal conductance at temperature $T$ \cite{Villagonzalo1999}.
Both electrons and phonons contribute to the heat current. When electron-phonon coupling is weak, the thermal conductance can be expressed as $\kappa=\kappa_\mathrm{el} + \kappa_\mathrm{ph}$. Enhanced thermoelectric efficiency requires simultaneous reductions in both contributions to thermal conductance without significantly affecting electronic conduction. However, this is not feasible in bulk metallic materials because the classical Wiedemann-Franz (WF) law states that the ratio $\mathcal{G} T/\kappa_\mathrm{el}$ is a universal constant \cite{Franz1853}.
Several strategies have been proposed to enhance the performance of thermoelectric materials. These include: (i) controlled reduction of lattice thermal conductance $\kappa_\mathrm{ph}$ by increasing phonon scattering, for example, via nanopatterning with antidots, defects, or edge modifications \cite{sadeghi2015enhanced,Gayner16,li2010high,shi2020advanced}; (ii) enhancement of electron-hole asymmetry at the Fermi energy, such as through edge roughness or the proximity effect of ferromagnetic insulator substrates \cite{ping2014valley,jia2021thermoelectric,ochi2023electron}; and (iii) modulation of diluted disorder potentials introduced by impurities, point defects, or vacancies \cite{an2014vacancy,PhysRevB.99.235428,gao2020thermoelectric,gupta2021first,jia2021thermoelectric,jha2021strategies,ngo2023effects}.

Recent studies have presented both theoretical~\cite{Hicks93,Khitun00,Balandin03,Sadeghi15} and experimental evidence~\cite{Venkata01,Harman02,Hochbaum08,Boukai08} indicating that nanostructuring enables thermoelectric efficiencies unattainable in bulk materials.
In these nanoscale systems, thermal and electronic properties can be independently tuned to achieve efficiencies that exceed the classical limit. Reported $ZT$ values surpass 2.4 in engineered nanostructured materials, including superlattices~\cite{Venkata01}, nanowires~\cite{Hochbaum2008}, and quantum dots~\cite{Harman2002}. Within this framework, silicene sheets and silicene nanoribbons (SNRs) are considered promising candidates for achieving high thermoelectric efficiencies
\cite{Zberecki2013,pan2012,nunez2020tuning}.

\begin{figure}[t]
    \centering
    \includegraphics[width=.95\linewidth]{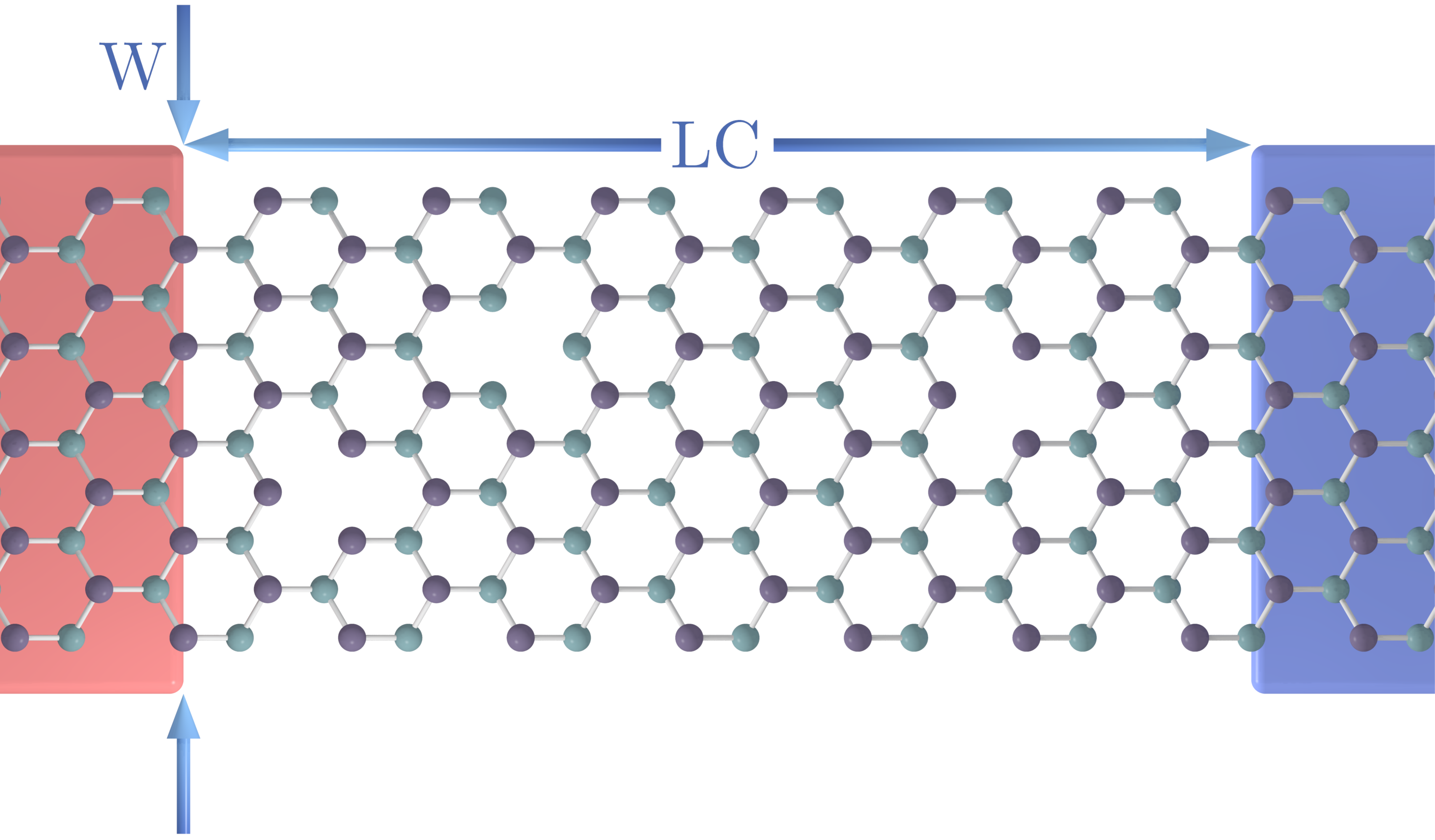}
    \caption{Schematic view of the proposed device. $LC$ and $W$ denote the length and width of the central ribbon, respectively. In this configuration, three atoms are removed, corresponding to a vacancy concentration of 3\%. Red and blue colors indicate the possible magnetic polarizations of the leads.}
    \label{SNR}
\end{figure}

Equilibrium molecular dynamics simulations indicate that bulk silicene exhibits an in-plane thermal conductivity of approximately $\unit{20}{\watt\per\metre\usk\kelvin}$ at room temperature \cite{Li2012}. This thermal conductivity is an order of magnitude lower than that of bulk silicon. Although silicene possesses a honeycomb lattice structure, the large ionic radius of silicon atoms induces lattice buckling \cite{Aufray10,Lalmi10}. This buckling modifies the vibrational modes of silicene, leading to both phonon softening and stiffening, thereby reducing phonon-mediated heat conduction.
Quantum effects allow thermoelectric devices to surpass the constraints of the classical Wiedemann-Franz (WF) law. Nanodevices exhibiting sharp resonances in electron transmission, such as those characterized by Fano lineshapes, are promising candidates for efficient heat-to-electricity conversion. In these systems, the ratio $\mathcal{G} T/\kappa_\mathrm{el}$ can significantly exceed the classical WF limit~\cite{Mahan96,GomezSilva12,Zheng12,Garcia13,Fu15,SaizBretin16,Wang16}.

In this work, we investigate the spin-dependent thermoelectric properties of a SNR heterostructure composed of a central conductor with a random distribution of vacancies, connected to two pristine leads of the same material placed over ferromagnetic insulators. We consider that the magnetic moments of the left and right leads are either parallel or antiparallel. A tight-binding Hamiltonian within the Green’s function formalism and the linear approximation is employed to calculate the spin-resolved electronic conductance, the charge and spin-dependent Seebeck coefficients, the total thermal conductance (including both phonon and spin-resolved electronic contributions), and the charge and spin figures of merit as functions of geometric confinement and vacancy concentration.
The results indicate an enhancement in both charge- and spin-dependent thermopower, resulting in improved thermoelectric efficiency at room temperature.

\section{\label{sec:theory} Model}

The system under study consists of a rectangular armchair-edged SNR (A-SNR) of width $W$ (measured in terms of the number of transverse Si-dimmers of the ribbon, $N$) and length $LC$ (measured in terms of the unit cell of the ribbon) connected to the source and drain leads, as shown schematically in Fig.~\ref{SNR}.  Electronic transport in these ribbons is described using a single-band tight-binding approximation \cite{Ezawa12a}. 
For this purpose, the system is divided into three spatial regions: the left contact, the scattering region (conductor), and the right contact. Thus, the Hamiltonian of the A-SNR can be written in the single-band tight-binding scheme as \cite{Ezawa12a}: 
\begin{equation}
\begin{aligned}
H ={}& -t \sum_{\langle i,j \rangle \sigma} c_{i\sigma}^\dagger c_{j\sigma}
+ i \frac{\lambda_{\mathrm{SO}}}{3\sqrt{3}} 
\sum_{\langle\langle i,j \rangle\rangle \sigma \beta} 
\nu_{ij} c_{i\sigma}^\dagger \sigma_z c_{j\beta} \\
&+ \sum_{i\sigma} c_{i\sigma}^\dagger (M \sigma_z + \epsilon_i) c_{i\sigma} \ ,
\label{Hamil}
\end{aligned}
\end{equation}
where $c_{i,\sigma}$ $(c_{i,\sigma}^{\dagger})$ annihilates (creates) an electron at $i$-th site. Sums over $\langle i, j \rangle$ and $\langle\!\langle i, j \rangle\!\rangle$ run over nearest and next-nearest neighbor sites, respectively. The spin indices $\uparrow$, $\downarrow$ are indicated by $\sigma$ and $\beta$ hereafter. The first term
of the Hamiltonian in Eq.\,(\ref{Hamil}) corresponds to nearest-neighbor hopping with energy $t = 1.6~\mathrm{eV}$. 
The second term represents the spin-orbit coupling with $\lambda_{\mathrm{SO}} = 3.9~\mathrm{meV}$, 
where $\nu_{ij} = \pm 1$ is the Haldane factor and $\sigma_z$ is the Pauli spin matrix. 
Finally, the third term takes into account the Zeeman spin-splitting of electron states due to the interaction with the magnetic substrate. For simplicity, we assume that the spin-splitting 
is the same in both sub-lattices.

We have considered a random distribution of vacancies, of concentration $C$, in the central region of a semiconductor ribbon, which are represented as missing atoms in Fig.~\ref{SNR}. From a numerical perspective, for calculating electronic properties, vacancies are simulated by setting a large on-site energy, approximately $10^4$ times the A-SNR energy scale, thereby preventing electrons from occupying those sites. On the other hand, for the phonon thermal conductance calculation, we have simulated the vacancies by removing atoms in the lattice, setting zero the inter-atomic potential between the vacancy and the nearest-neighbor atoms.

\begin{figure*}[t]  
  \centering
  \subfigure{\includegraphics[width=0.48\textwidth]{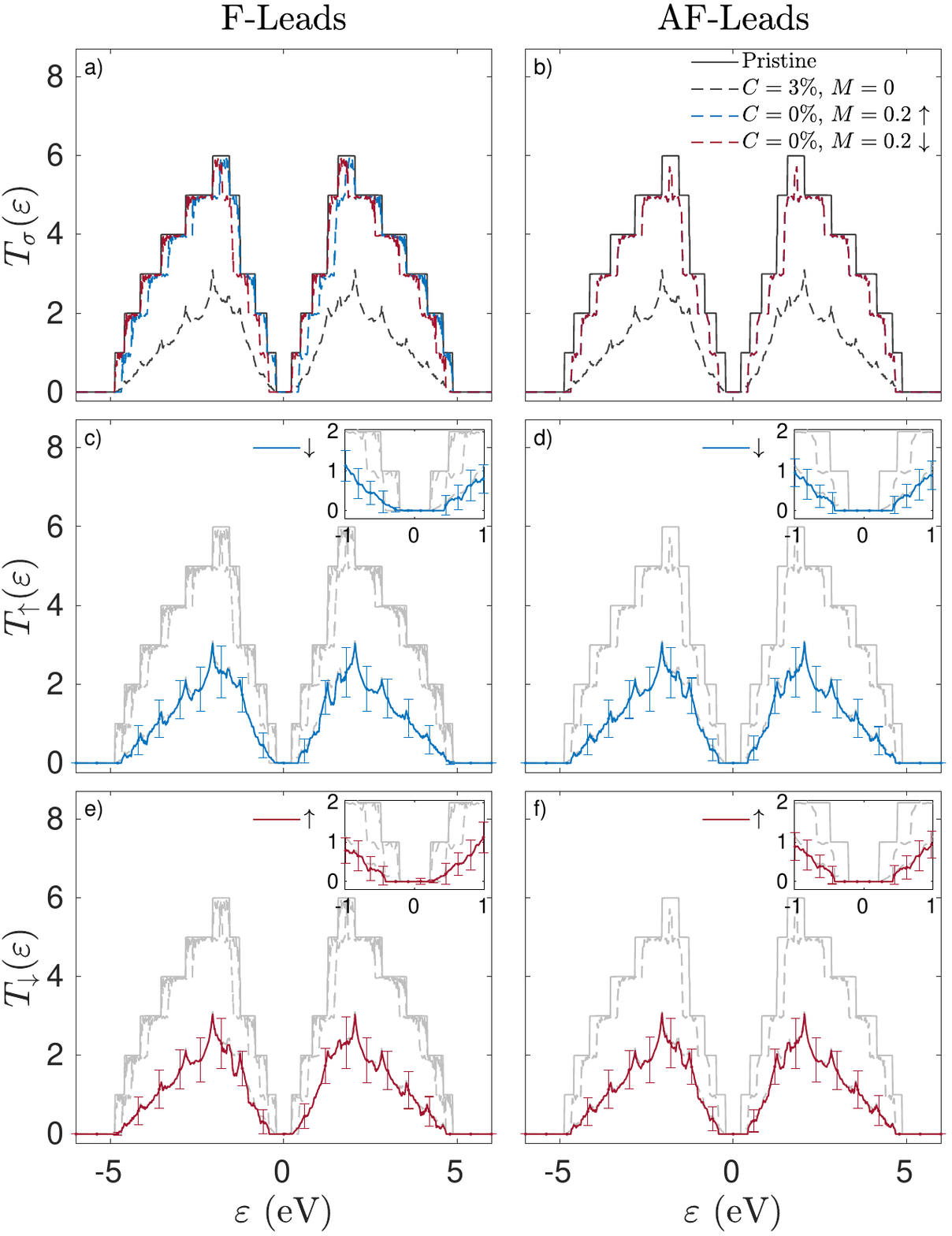}}
  \hspace{0.02\textwidth}
  \subfigure{\includegraphics[width=0.48\textwidth]{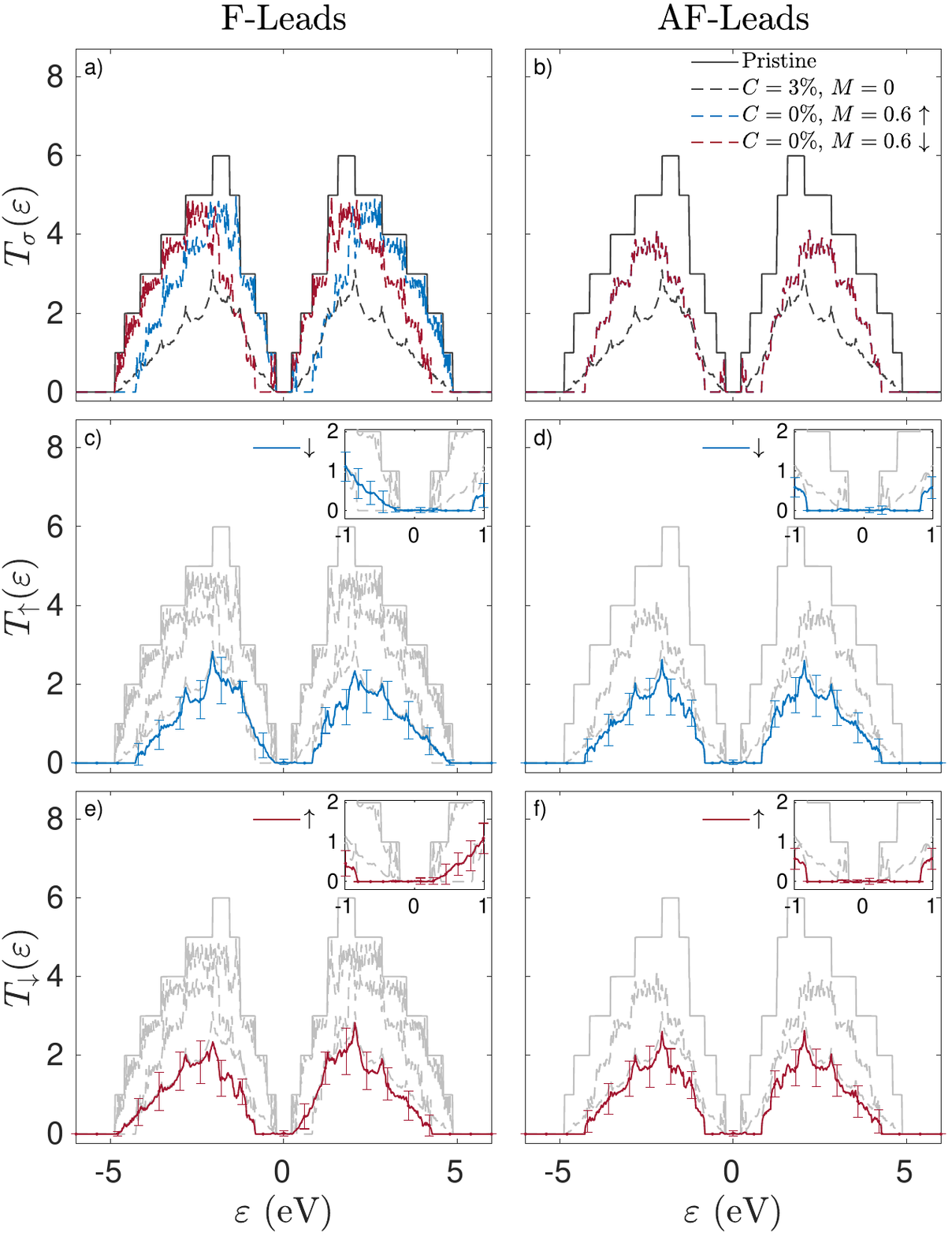}}
  \caption{Spin-dependent transmission $T_\sigma(\varepsilon)$ for ferromagnetic and antiferromagnetic configurations in A-SNRs with $N=12$, $LC=20$ ($L \approx 26$~nm) and $C=3\%$. Left panels: $M=0.2$ eV; right panels: $M=0.6$ eV. In both cases, systems with no vacancies (0\%) and with 14 vacancies (3\%) are considered.}
  \label{trM02}
\end{figure*}

To analyze the thermoelectric behavior of the defective A-SNR, we adopt the linear response approximation, where an effective voltage drop, $\Delta V$, and a temperature difference, $\Delta T$, are applied between the left and right contacts.
Within this approach, the spin-resolved electronic current $I_{\text{e},\sigma}$ and the heat current $I_{\text{Q},\sigma}$ are:
\begin{eqnarray}
I_{\text{e},\sigma} &=& - e^2 L_{0,\sigma} \Delta V + \frac{e}{T}\, L_{1,\sigma} \Delta T\ , \label{Ie} \\
I_{\text{Q},\sigma} &=& e L_{1,\sigma} \Delta V - \frac{1}{T}\, L_{2,\sigma} \Delta T\ , \label{Iq}
\end{eqnarray}
where $e$ is the electron charge, $T$ is the absolute temperature, and $L_{n,\sigma}$ (with $n= 0,1,2$, and $\sigma$ is the spin index) are thermal integrals defined as:
\begin{equation}
    L_{n,\sigma} (\mu, T) = \frac {1} {h} \int_0 ^ \infty dE \, \mathcal {T}_{\sigma} (E)\; (E- \mu) ^ n \left (- \frac {\partial f (E, T)} {\partial  E} \right)\ .
    \label{Ln}
\end{equation}
Here, $h$ is Planck's constant, $\mu$ is the chemical potential, $f(E,T)$ is the equilibrium Fermi-Dirac distribution, and $\mathcal{T}_\sigma(E)$ is the energy- and spin-dependent electronic transmission probability.

The electronic conductance is defined as $\mathcal{G}_{\sigma} = -I_{\text{e}}/\Delta V$ and it can be obtained directly from Eq.~(\ref{Ie}) as
\begin{equation}
    \mathcal{G}_{\sigma}(\mu,T) = e ^ 2 L_{0,\sigma} .
    \label{Conductance}
\end{equation}
The Seebeck coefficient $S$ is calculated in the linear response regime, namely, $| \Delta T | \ll T$ and $| e \Delta V | \ll \mu$. It is defined as the voltage drop induced by a temperature gradient at zero electric current, $S = \Delta V / \Delta T|_{I_{\text{e}}} = 0$ [see Eq.~(\ref{Ie})], in the limit $\Delta T \to 0$. Thus,
\begin{equation}
    S_{\sigma}(\mu,T) = - \frac {1} {e T} \frac {L_{1,\sigma}} {L_{0,\sigma}} \ .
    \label{Seebeck}
\end{equation}
The electronic contribution to the thermal conductance is defined as the ratio between the thermal current $I_Q$ and the temperature gradient $\Delta T$ when the electric current $I_e$ is zero $\kappa_\mathrm{el} = I_Q / \Delta T|_ { I_e = 0}$. Written in terms of the thermal integrals, it is given by
\begin{equation}
    \kappa_\mathrm{el,\sigma} (\mu, T) = \frac {1}{T}\left(L_{2,\sigma}-\frac{L_{1,\sigma}^2}{L_{0,\sigma}}\right)\ .
    \label{ThermalConductanceE}
\end{equation}
The total thermal conductance of the SNRs is obtained as $\kappa(\mu, T) = \kappa_\mathrm{el}(\mu, T) + \kappa_\mathrm{ph}(T)$. It should be emphasized that $\kappa_\mathrm{ph}(T)$ depends only on temperature and not on the chemical potential $\mu$. 
For thermoelectric quantities, particularly the thermal conductance of phonons, $\kappa_\text{ph}(T)$, we used QuantumATK 2022.12 \cite{ATK} to compute it as a function of temperature, geometric parameters $N$ and $LC$, and various concentrations of vacancies $C$. The results were averaged over hundred of configurations representing the same type of disorder due to vacancies in the nanoribbon. In our study, we can neglect electron–phonon and phonon–phonon interactions \cite{Ouyang,
Gunlycke} within the conductor because these interactions are of higher order in comparison to the harmonic interaction term, which we have used to describe qualitatively ballistic thermal transport. In this approximation, phonon transport can be calculated similarly as the electronic counterpart. 

The thermoelectric efficiency is determined by the figure of merit, expressed as follows:
\begin{equation}
    ZT_{\sigma}(\mu,T) = \frac{\mathcal{G}_{\sigma}\,S_{\sigma}^2\,T}{\kappa_{\mathrm{el},\sigma}(\mu,T) + \kappa_\mathrm{ph}(T)} \ .
    \label{Seebeck2}
\end{equation}
For narrow silicene nanoribbons, it has been shown that the lattice component of the thermal conductivity is lower than the electronic contribution, mainly due to the abrupt reduction of the phonon mean free path. In particular, this holds for low-gap A-SNRs, which are the fundamental components of our system.
\cite{broido2007intrinsic,ward2009ab}.

Finally, we define the charge and spin electronic conductances as:
\begin{equation}
     \mathcal{G}_c = (\mathcal{G}_{\uparrow} + \mathcal{G}_{\downarrow})\text{ ,} \hspace{1cm} \mathcal{G}_s = (\mathcal{G}_{\uparrow} - \mathcal{G}_{\downarrow})\text{ ;}
\end{equation}
and the charge and spin thermoelectric quantities as follows: 
\begin{equation}\label{Scs}
     S_c = \frac{(S_{\uparrow} + S_{\downarrow})}{2}\text{ ,} \hspace{1cm} S_s = (S_{\uparrow} - S_{\downarrow})\text{ ;}
\end{equation}
\begin{equation}
     \kappa_{\text{el}} = (\kappa_{\text{el},\uparrow} + \kappa_{\text{el},\downarrow})\text{ ,} \hspace{1cm} \kappa_\text{tot} = (\kappa_{\text{el}} + \kappa_{\text{ph}})\text{ .}
\end{equation}
Accordingly, the charge and spin figure of merit are expressed as:
\begin{equation}\label{ZTcs}
     ZT_c = (ZT_{\uparrow} + ZT_{\downarrow})\text{ ,} \hspace{1cm} ZT_s = (ZT_{\uparrow} - ZT_{\downarrow})\text{ .}
\end{equation}

\section{Thermoelectric properties of SNRs}\label{sec:res}

In what follows, we focus on the spin-dependent thermoelectric response of A-SNRs as a function of geometric parameters, such as, the number of dimers $N$, the length of the nanoribbon $LC$, the concentration of vacancies $C$, and the magnetic exchange coupling $M$ in both leads.  

Figure~\ref{trM02} presents the spin-dependent transmission probability $\mathcal{T}_{\sigma}(E)$ for a semiconductor A-SNR with length $L = 26$ nm, width $N = 12$, and a vacancy concentration of 3\%. For comparison, the corresponding pristine A-SNR connected to magnetic leads is also included. The spin-resolved transmission is shown for magnetic exchange couplings $M = 0.2$ eV (left panels) and $M = 0.6$ eV (right panels), considering both ferromagnetic and antiferromagnetic configurations of the leads.
Panels (a) and (b) in both groups of plots compare the effects of magnetic contacts in a pristine A-SNR ($C=0\%$; $M=0.2$ eV and $M=0.6$ eV) with the effects of vacancies in an A-SNR without magnetic contacts ($C=3\%$; $M=0$ eV). This comparison facilitates visualization of the competition between these two phenomena and enables analysis of their impact on the transport properties of the system.

Panels (a) and (b) show that, for both values of $M$, the ferromagnetic configuration exhibits the characteristic Zeeman spin splitting of the transmission across all transverse channels, particularly near the Fermi level, where electron-hole asymmetry emerges. In contrast, the antiferromagnetic configuration does not display spin splitting; however, the transmission is more significantly affected than in the ferromagnetic case. This increased impact arises because the exchange field induces a spin-dependent shift in the energy levels of both contacts, thereby influencing the system's ballistic transport. As $M$ increases, transmission in both magnetic configurations is further suppressed, primarily due to Fabry–Pérot–like resonance effects that disrupt ballistic electronic transport in the ribbon. In the absence of magnetic leads, the introduction of a $3\%$ diluted vacancy concentration exerts a more pronounced effect on transmission than contact magnetization, even at high $M$ values. This is attributed to vacancies causing electron scattering, which reduces the mean free path and inhibits the ballistic transport regime, as previously reported \cite{nunez2020tuning}.
Panels (c) to (f) present the spin-up and spin-down transmission probabilities at two exchange field values ($M=0.2$ eV and $M=0.6$ eV), considering a $3\% $
vacancy concentration randomly distributed within the ribbon. The results represent averages over hundreds of random realizations. As previously discussed, spin-resolved transmission is significantly influenced by the presence of defects. Notably, a distinct difference between the two magnetic configurations emerges, associated with electron-hole asymmetry near the Fermi energy. This asymmetry is pronounced only in the ferromagnetic configuration (as shown in the insets) and becomes more prominent with increasing magnetic polarization. This phenomenon will govern the spin-dependent thermoelectric response of the ribbon.

\begin{figure}[th]
    \centering
    \includegraphics[width=1.0\linewidth]{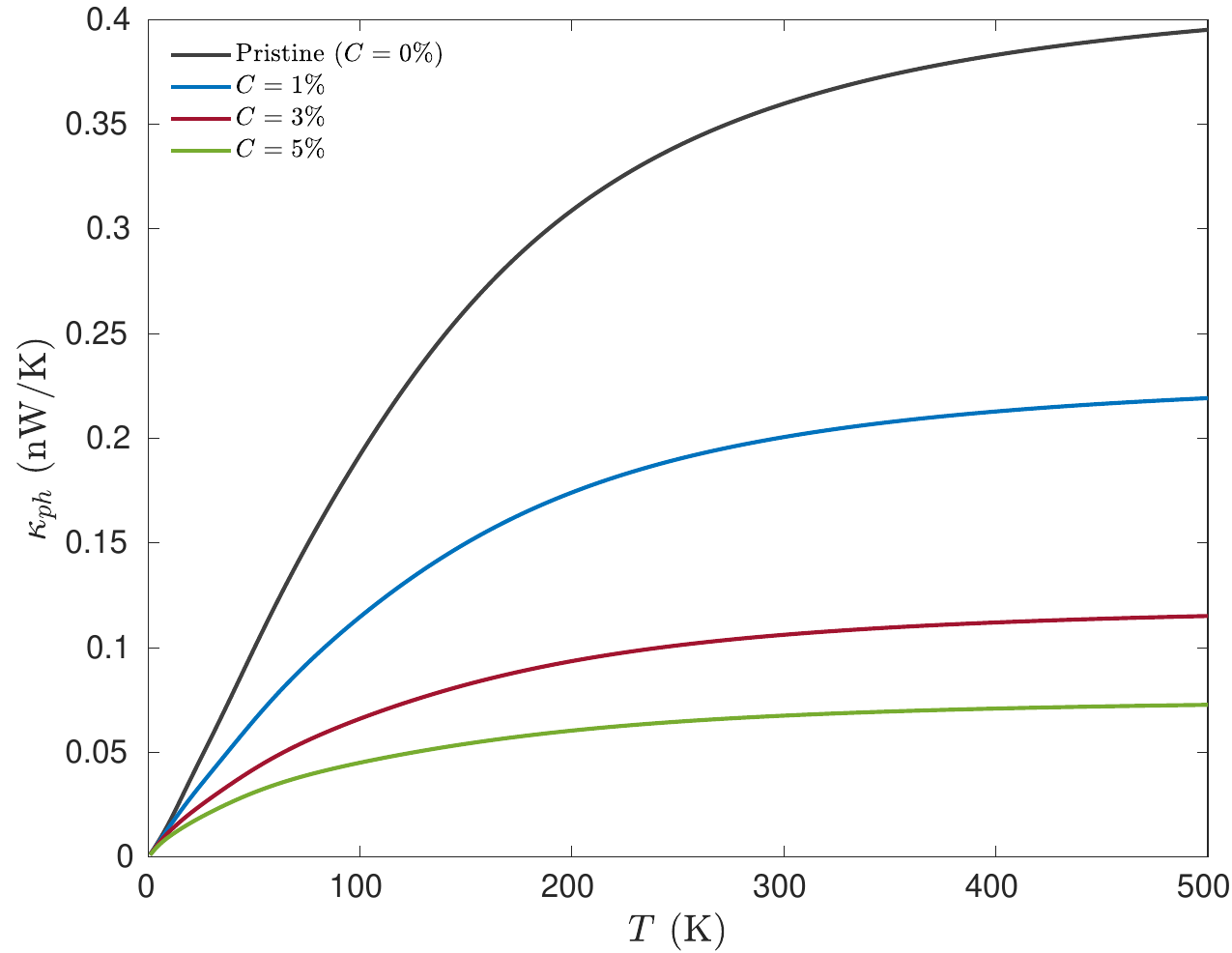}
    \caption{Phonon thermal conductance $\kappa_\text{ph}$ as a function of temperature for $N=12$, $L=26$ nm, computed with QuantumATK 2022.12 \cite{ATK}.}
    \label{ph_th_cond}
\end{figure}

\begin{figure}[th]
\centering
\includegraphics[width=1.0\linewidth]{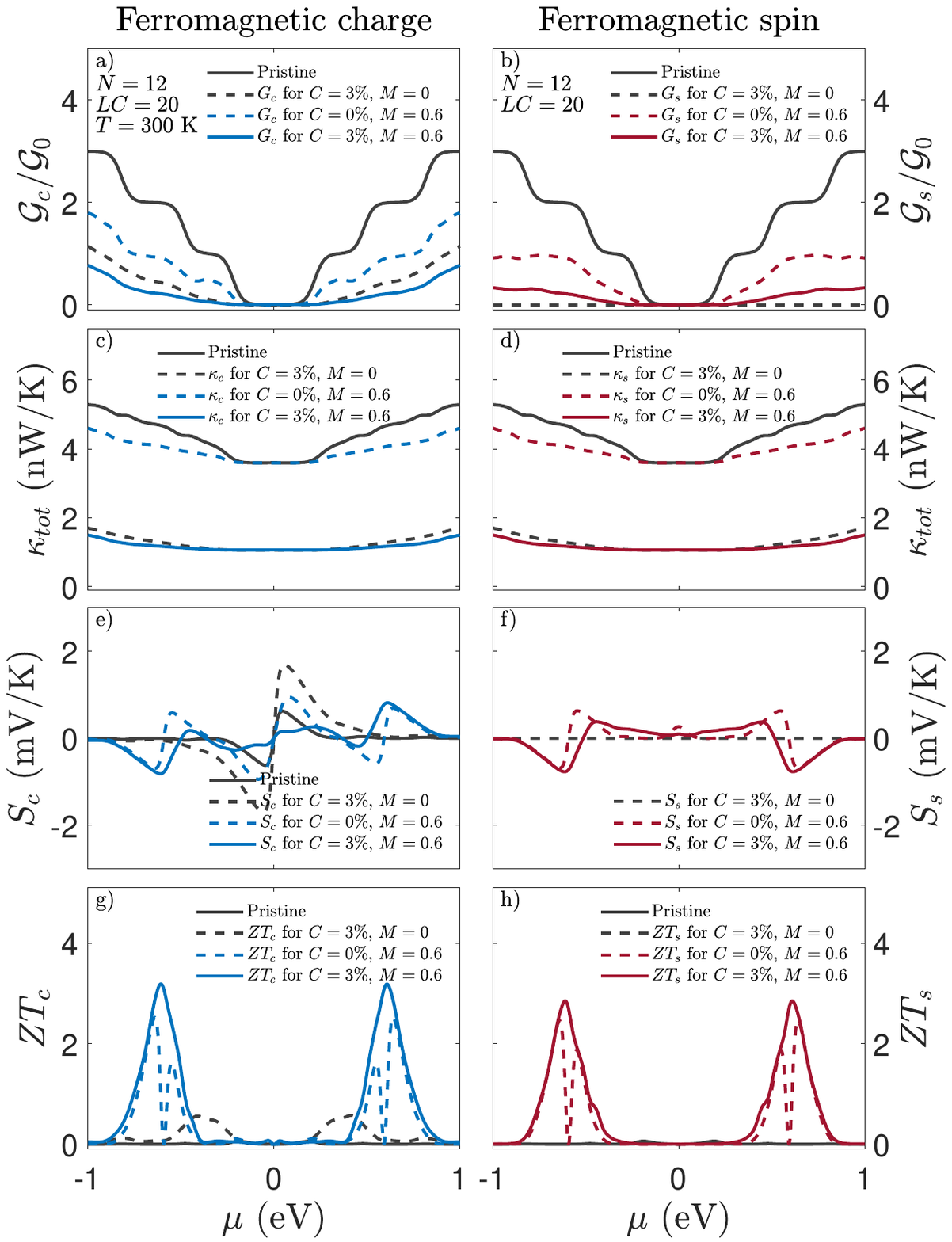}
\caption{[(a) and (b)] Electronic conductance $\mathcal{G}_{c,s}$ in units of $\mathcal{G}_{0}=e^2/h$, [(c) and (d)] total thermal conductance $\kappa_{\text{tot}}$, [(e) and (f)] Seebeck coefficient $S_{c,s}$, and [(g) and (h)] figure of merit $ZT_{c,s}$ for $N=12$, $L\approx 26$ nm, 0$\%$ and 3$\%$ vacancies, and $M=0.6$ for the ferromagnetic configuration.}
\label{th_N12}
\end{figure}
We now examine how the phonon thermal conductance is modified by the presence of vacancies. 
Figure~\ref{ph_th_cond} displays $\kappa_{\mathrm{ph}}$ as a function of temperature for a semiconductor A-SNR with width $N = 12$, length $L = 26$ nm, and varying vacancy concentrations ($C = 0\%, 1\%, 3\%,$ and $5\%$). For the pristine A-SNR, at low temperatures ($T < 100$ K), $\kappa_{\mathrm{ph}}$ is approximately 0.18 nW/K, primarily determined by longitudinal and transverse acoustic phonon modes. As temperature increases and the system remains in the ballistic regime, optical phonon modes become progressively activated, leading the thermal conductance to approach saturation. It should be noted that anharmonic effects, including phonon–phonon and electron–phonon interactions, are intrinsic. However, at temperatures below the Debye temperature ($\Theta_{S_i} = 640$ K), these contributions are weak and may be treated as second-order perturbations.

Vacancy defects modify the electronic properties of materials and serve as scattering centers that induce both phonon localization and delocalization, which reduces thermal transport \cite{hao, Wirth}. Such defects can substantially suppress phonon transport in graphene \cite{Zhao2011}, silicene nanowires \cite{Hu}, and silicene nanoribbons (SNRs) \cite{Pan}. For instance, a single vacancy in a 400-atom silicene sheet can decrease thermal conductance by approximately 78\% \cite{Li2012}, underscoring the critical role of vacancies in regulating phonon transport at the nanoscale.
In this context, we have focused on the effects of a random distribution of single vacancies, in a diluted regime, on the phonon thermal transport of A-SNRs. 
Figure~\ref{ph_th_cond} demonstrates that $\kappa_{\mathrm{ph}}$ is substantially reduced across the entire temperature range for all vacancy concentrations, primarily due to defect–phonon scattering. At low temperatures, long-wavelength acoustic phonons dominate thermal transport, resulting in $\kappa_{\mathrm{ph}}$ values that remain lower than those of pristine samples. While the temperature dependence follows a similar trend, vacancy concentrations exceeding 1\% reduce $\kappa_{\mathrm{ph}}$ by more than 60\%, with saturation observed near 300 K. These findings indicate that vacancies reduce the mean free path of both optical and acoustic phonons.
This behavior arises from the vibrational dynamics of the sub-lattices. In acoustic modes, the sub-lattices oscillate nearly in phase over regions smaller than the phonon wavelength, resulting in reduced sensitivity at low energies. In contrast, in optical modes, the sub-lattices vibrate out of phase, which causes even long-wavelength optical phonons to be influenced by disorder at the scale of the lattice parameter \cite{nunez2020tuning}.

The analysis now turns to the thermoelectric response of the semiconductor A-SNR in the presence of a random vacancy distribution. Using the definitions provided in Eqs.~(\ref{Conductance})–(\ref{ZTcs}), the charge and spin electronic conductance $\mathcal{G}_\sigma$, the charge and spin Seebeck coefficients $S_\sigma$, the total thermal conductance $\kappa_{\mathrm{tot},\sigma}$, and the charge and spin figures of merit $ZT_\sigma$ are computed.
In the antiferromagnetic configuration, $G_{_S} \sim0$, $S_{_S} \sim0$, and $ZT_{_S} \sim0$ as a result of the symmetry of the spin-split bands in the contacts and the resulting absence of electron–hole asymmetry around the Fermi level.
Figure~\ref{th_N12} displays the average thermoelectric properties of a semiconductor A-SNR with width $N = 12$, length $L = 26$ nm, and vacancy concentrations $C = 0\%$ and $3\%$, for the ferromagnetic configuration at a temperature $T = 300$\,K. For reference, the corresponding pristine A-SNR is also shown (black solid curve).

\begin{figure*}[th]
  \centering
  \subfigure{\includegraphics[width=0.48\textwidth]{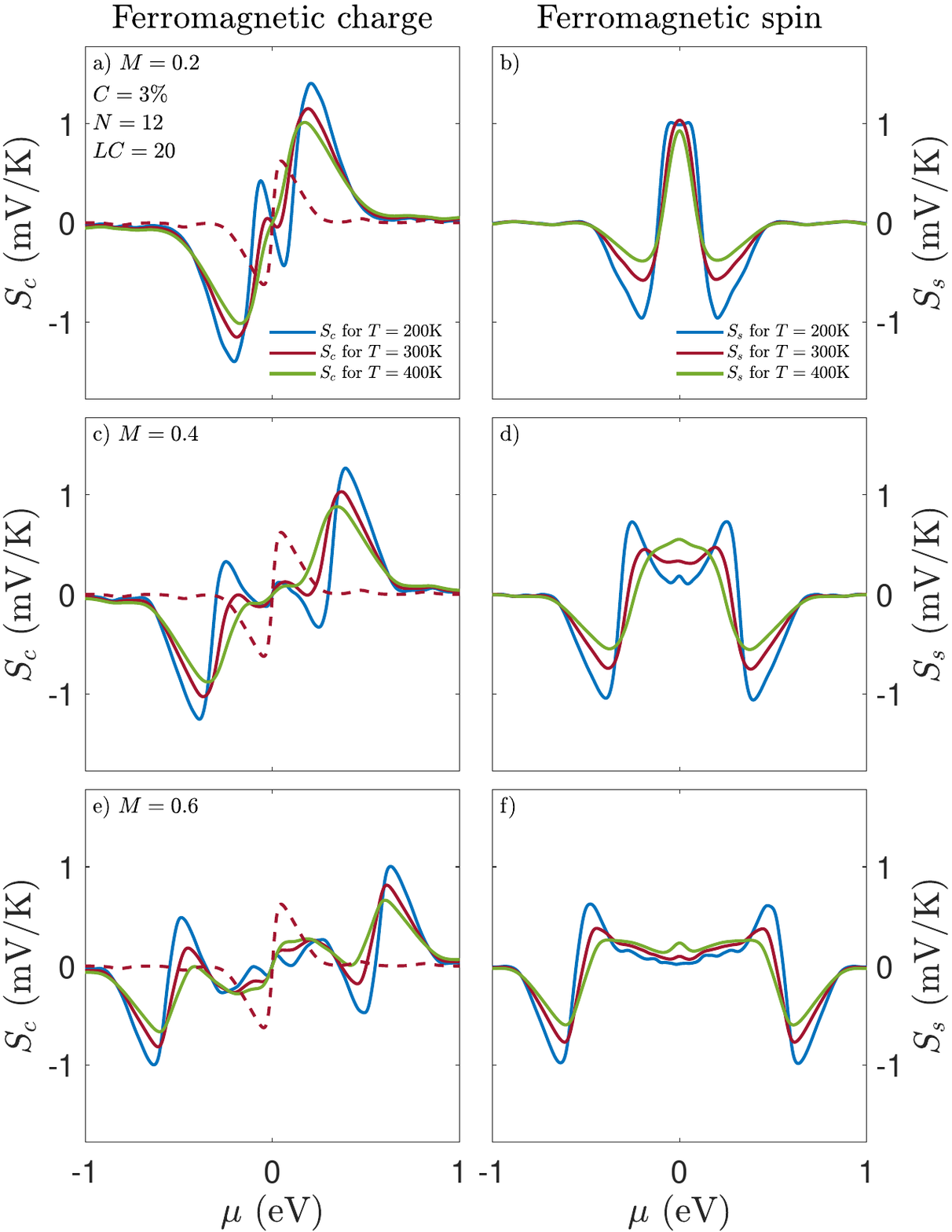}}
  \hspace{0.02\textwidth}
  \subfigure{\includegraphics[width=0.48\textwidth]{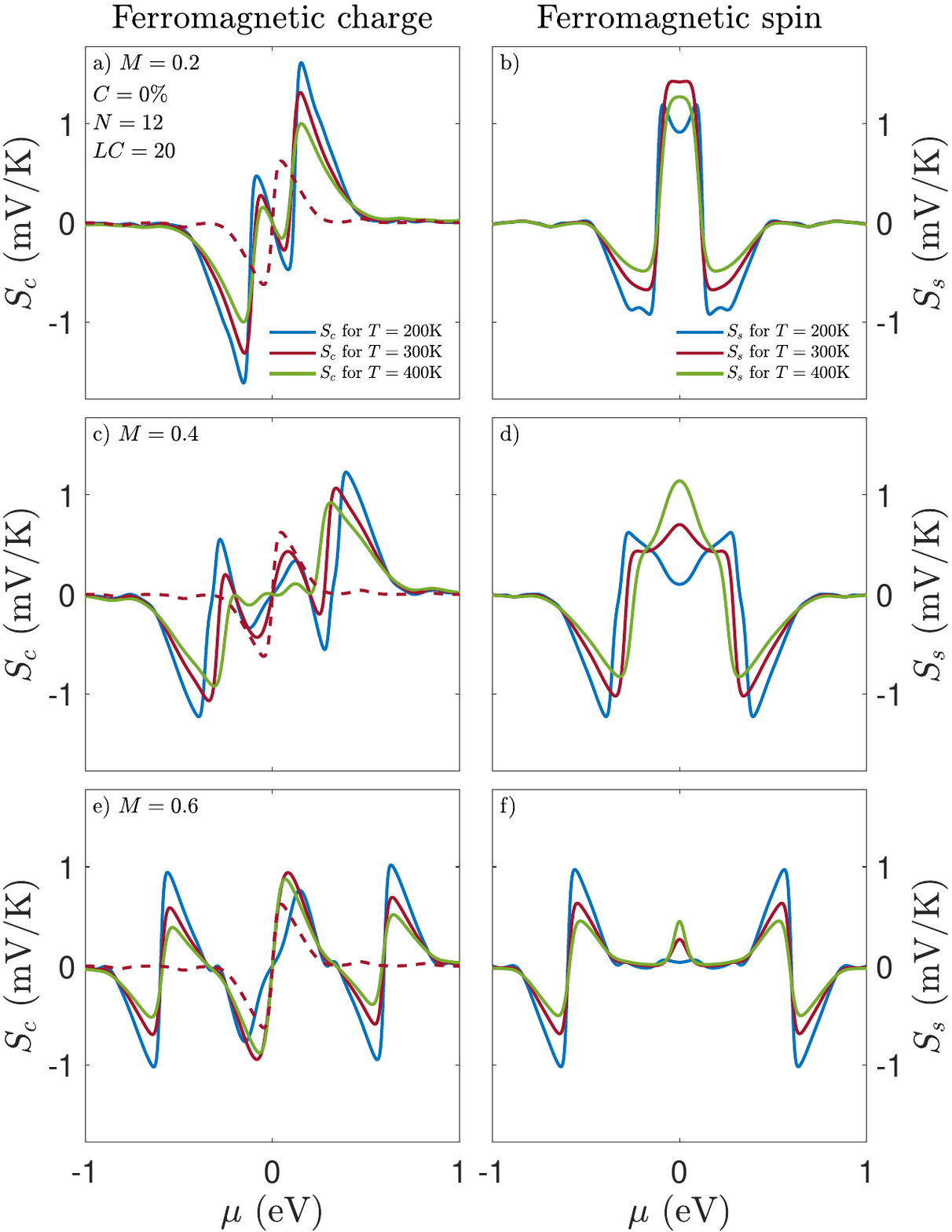}}
  \hspace{0.02\textwidth}
   \caption{Comparison of the charge and spin Seebeck coefficients for an  N=12 A-SNR. Left panels correspond to the system with diluted disorder of 3\% and ferromagnetic contacts (for different exchange field intensities), while right panels show the defect-free case with ferromagnetic leads. The dashed red curves correspond to charge Seebeck to the pristine case, with $M=0$ and $C=0$ at $T = 300 $ K.}
   \label{compS} 
\end{figure*}

Analysis of the charge and spin conductance curves indicates that the combined presence of ferromagnetic contacts and a 3\% vacancy concentration significantly degrades the  transport properties of the system. Although this suppression is substantial, finite charge and spin transport persist. The disorder potential introduced by vacancies can be interpreted within the one-parameter scaling theory of disordered systems, as introduced by Abrahams et al. \cite{Abrahams}. According to this framework, conductance is exponentially suppressed with increasing system length, described by $G(l) \sim e^{-\gamma l}$, where $\gamma$ denotes the decay rate. This relationship enables the definition of a localization length, $\lambda_\sigma = 1/\gamma$, which depends on spin direction, energy, and vacancy concentration. In short ribbons, even relatively weak and dilute disorder can induce a conductance suppression exceeding 70\% \cite{Nuñez_2016}. Beyond localization effects, ferromagnetic leads further degrade conductance by forming an effective double potential barrier for electrons, resulting in a Fabry–Pérot–like interference pattern that restricts electronic transport through the conductor.
In the case of the total thermal conductance curves, it can be observed that the maximum deterioration is achieved for the combined presence of a 3\% vacancy concentration and ferromagnetic contacts with an exchange splitting of $M=0.6$ eV. This result can be understood as a direct consequence of the strong scattering introduced by vacancies, which leads to a significant reduction of the mean free path of both acoustic and optical phonons. Consequently, phonon transport across the ribbon is severely limited, giving rise to a pronounced suppression of the total thermal conductance. 

Given that both total thermal conductance and electronic conductance decrease due to the combined effects of vacancies and ferromagnetic contacts, it is essential to analyze thermopower behavior. Because thermoelectric efficiency depends quadratically on thermopower, enhancements in the charge and spin Seebeck coefficients within specific energy ranges can significantly improve overall device performance. The results indicate that both spin and charge Seebeck coefficients are enhanced near energies of ±0.55 eV for an appropriate combination of exchange field strength and dilute disorder, while finite transport is maintained. This enhancement arises from two complementary mechanisms. First, according to the Sommerfeld expansion of the thermal integrals, the Seebeck coefficient is directly related to the ratio $\mathcal{T}'(\varepsilon)/\mathcal{T}(\varepsilon)$, indicating that it increases when the transmission function exhibits sharp energy-dependent variations. Such features are introduced by impurities in the ribbon, which cause strong energy-dependent scattering, as previously discussed in Fig. \ref{trM02}. Second, the Seebeck coefficient can be further increased by enhancing electron–hole asymmetry. In this system, ferromagnetic contacts naturally provide this asymmetry, contributing to the enhancement of $S_{c,\sigma}$. The combined suppression of thermal and electronic conductances, together with increased thermopower, results in improved charge and spin figures of merit. Specifically, panels (g) and (h) of Fig.~\ref{th_N12} demonstrate that a 3\% vacancy concentration combined with ferromagnetic contacts ($M=0.6$ eV) enhances thermoelectric efficiency for both charge and spin transport compared to the pristine case with magnetic contacts.

\begin{figure*}[th]
  \centering
  \subfigure{\includegraphics[width=0.48\textwidth]{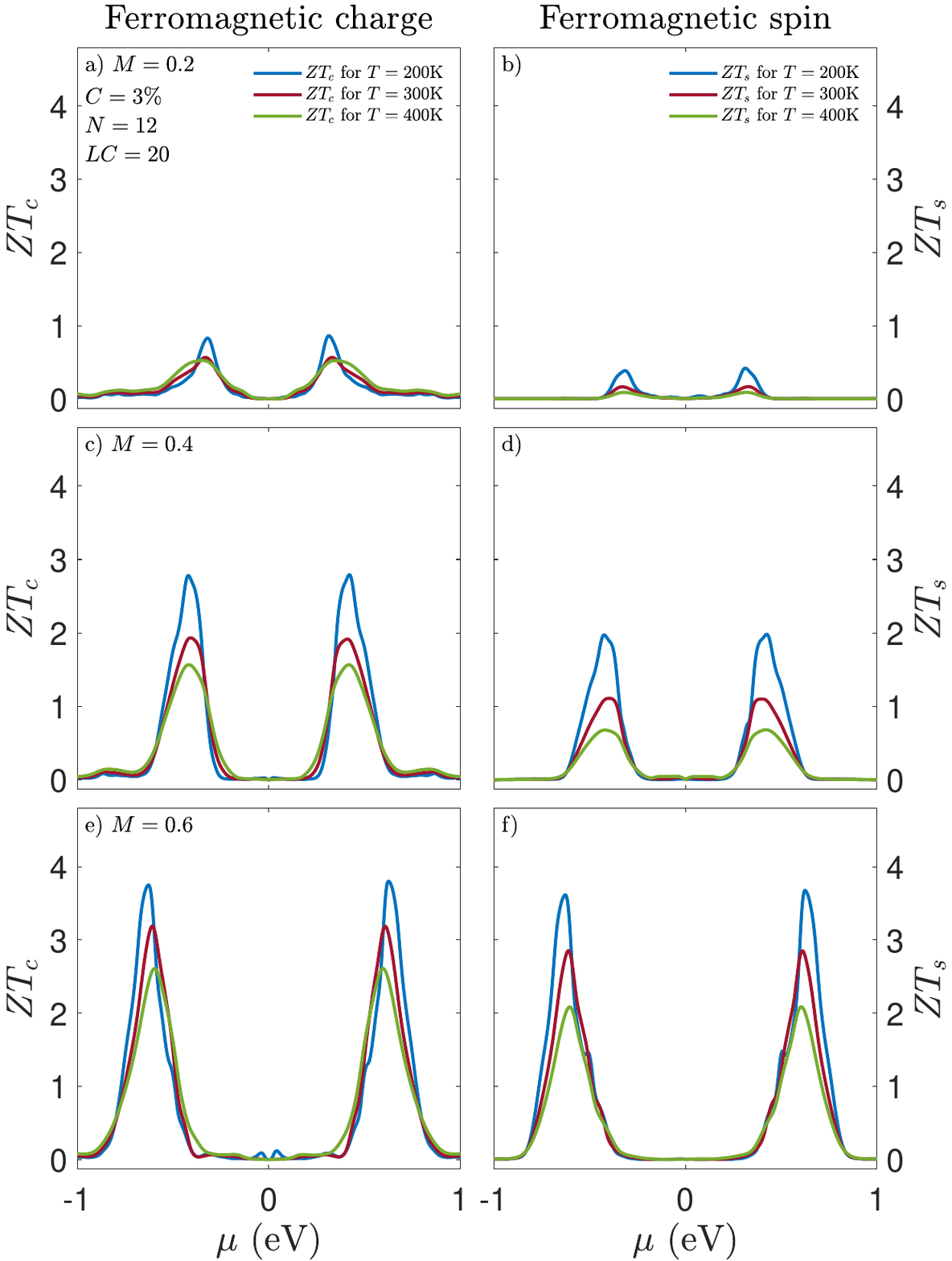}}
  \hspace{0.02\textwidth}
  \subfigure{\includegraphics[width=0.48\textwidth]{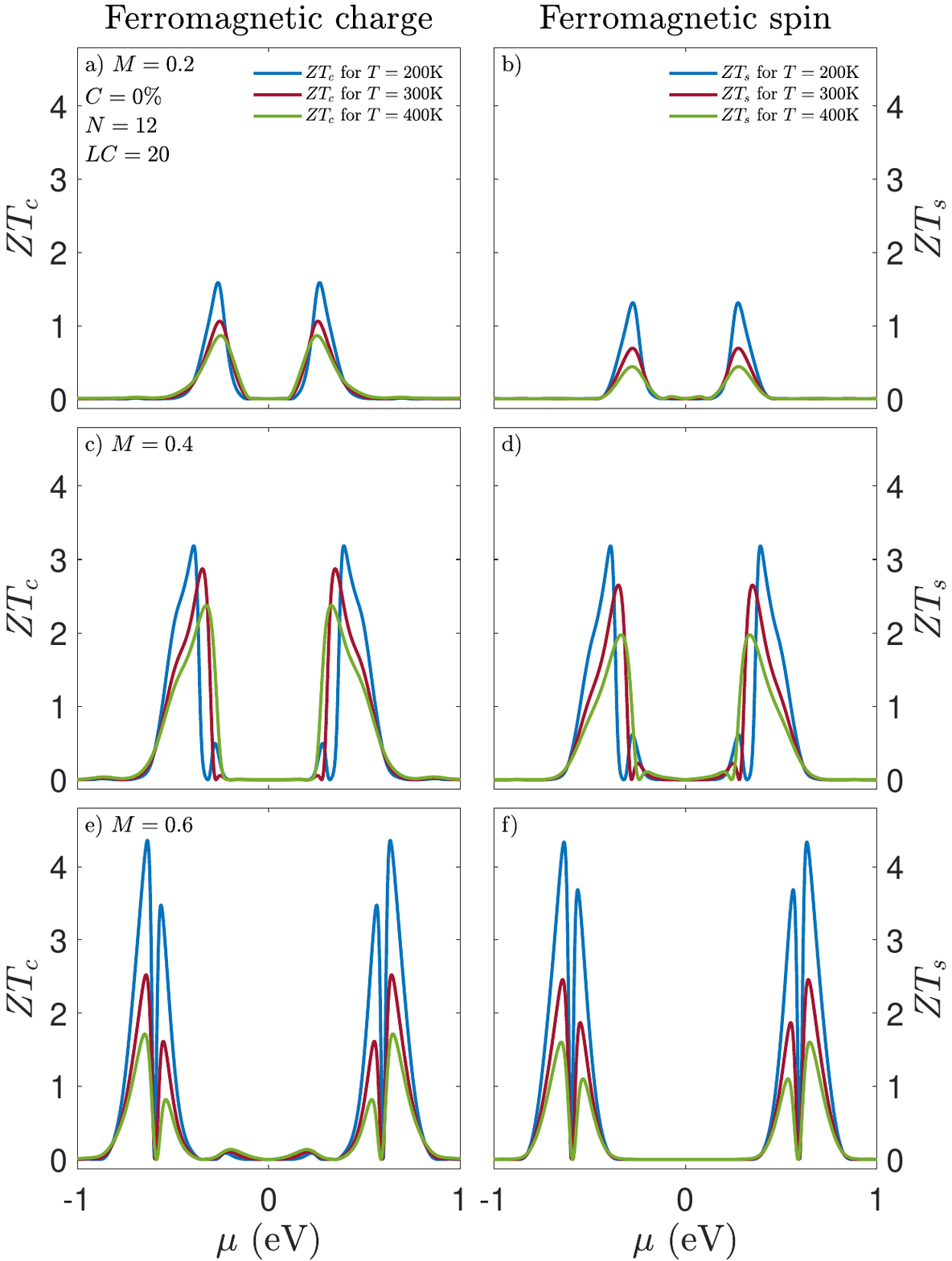}}
  \hspace{0.02\textwidth}
   \caption{Comparison of the charge and spin figure of merit for an  N=12 A-SNR. Left panels correspond to the system with diluted disorder of 3\% and ferromagnetic contacts (for different exchange field intensities), while right panels show the defect-free case with ferromagnetic leads.}
   \label{compZT}
\end{figure*}

To assess the robustness of the results displayed in Fig.~\ref{th_N12}, we calculated and compared the charge and spin Seebeck coefficients (Fig.~\ref{compS}) as well as the charge and spin figures of merit (Fig.~\ref{compZT}) for the $N=12$ A-SNR system with ferromagnetic leads. This analysis considered three distinct values of the exchange splitting, $M$, and three different temperatures. Comparisons were performed for both the disorder-free system and the system with the optimal disorder concentration (3\%).
Figure~\ref{compS} shows that, for the three values of the exchange field, the charge Seebeck coefficient in both systems displays a similar overall temperature dependence, with the maximum value decreasing as temperature increases. The most significant differences arise from pronounced electron–hole asymmetry induced by the combined effects of ferromagnetic contacts and impurity disorder. Specifically, impurities strongly suppress the thermopower near zero energy, and this suppression intensifies as $M$ increases. This phenomenon results from the reduction and smoothing of the electronic transmission curves around the Fermi energy due to disorder potentials, which consequently decrease the charge Seebeck coefficient in this energy range.
The spin Seebeck coefficient exhibits more pronounced differences between the pristine and disordered cases. Notably, due to the definition adopted in Eq.~(\ref{Scs}), electron–hole asymmetries are absent because they are compensated by the difference between $S_\uparrow$ and $S_\downarrow$. Across all values of the exchange field and temperatures, the overall magnitude of the spin Seebeck response in the pristine system remains consistently larger than in the disordered case. The most significant distinction between the corresponding curves is the strong suppression of the spin Seebeck coefficient over a wide energy range around the Fermi energy in the presence of disorder. This suppression is directly related to the transmission properties at these energies, where the difference between the two spin components is less pronounced than in the pristine system [see Fig. \ref{trM02}]. Consequently, the energy derivative of the transmission in the disordered case is smoother than in the pristine case, resulting in an almost vanishing spin Seebeck response at those energies.

The comparative analysis of the charge and spin figures of merit is presented in Fig. \ref{compZT}. Clear trends in the thermoelectric performance for both charge and spin transport can be identified. For low values of the exchange field ($M<0.4$ eV) and for all considered temperatures, both figures of merit are higher in the pristine ribbon. This behavior is primarily related to the more favorable charge and spin Seebeck responses observedhe defect-free system within this range of exchange fields.
However, when an optimal combination of parameters is achieved, specifically an exchange field of
$M=0.6$ eV combined with a vacancy concentration of
3\%, and at temperatures above 300 K, the charge and spin thermoelectric efficiencies increase in the disordered system. This enhancement originates from the strong electron–hole asymmetry induced by the exchange field, combined with an optimal reduction of the charge and thermal conductances. As a result, the figure of merit can be optimized in a controlled manner, highlighting the constructive role of a moderate amount of disorder and magnetic leads in improving the thermoelectric performance of the system.

\section{Final Remarks}\label{sec:con}

In this work,we have systematically investigated the thermoelectric properties of armchair silicene nanoribbons in the presence of diluted disorder and magnetic contacts, with particular emphasis on both charge and spin transport. Our analysis shows that the interplay between magnetic polarization and controlled disorder strongly influences the thermoelectric efficiency. In particular, an optimal combination of a finite exchange field $M$ in a ferromagnetic configuration, and a low concentration of vacancies leads to a controlled enhancement of the thermoelectric figure of merit.

In the absence of disorder, the pristine system with magnetic contacts demonstrates significantly improved thermoelectric performance across a wide range of parameter space compared to the pristine case without magnetic contacts. This enhancement is primarily attributed to the strong electron–hole asymmetry induced by the magnetic contacts, which increases the Seebeck response without substantially increasing the electronic and thermal conductances. The introduction of a moderate amount of disorder further reinforces this effect. Diluted defects optimally reduce charge and thermal conductances while maintaining a sizable thermopower, thereby further increasing thermoelectric efficiency.

In the antiferromagnetic configuration, the spin-dependent thermoelectric quantities remain nearly zero. This outcome results from the lack of significant electron–hole asymmetry in the spin-resolved electronic transmission curves, which suppresses both the spin Seebeck response and the spin-dependent electronic conductance. Although finite thermoelectric efficiencies are observed for the charge-related quantities, their absolute values remain consistently low and do not approach unity within the examined parameter space. Therefore, the antiferromagnetic configuration does not provide any thermoelectric advantage compared to the ferromagnetic case for any of the exchange field strengths, disorder concentrations, or temperatures considered in this work.

We have confirmed that these trends are not limited to a specific system size. Comparable behaviors were observed for multiple conductor lengths and ribbon widths, demonstrating that the principal conclusions of this study remain robust under geometric variation. Collectively, these results establish a reliable approach to improving the thermoelectric efficiency of silicene nanoribbon-based systems by integrating magnetic contacts with controlled disorder.

\acknowledgments

D.Z. acknowledges support from USM-Chile under Grant PI-LIR-2022-13. P.A.O. acknowledges support from FONDECYT Grants No. 1220700 and  1230933. J.P.R.-A is grateful for the financial support of FONDECYT Iniciaci\'on grant No. 11240637. L.R. thanks the financial support by the FONDECYT Grant No.1220700 and the DGIIP-USM grant.

\bibliography{references.bib} 

\end{document}